%
%
\def\BJBa4{
 \hsize=6.3in
 \vsize=9.6in
 \voffset=-0.3in
}

\def\laeq{\lower.5ex\hbox{\fiverm{\ $\buildrel < \over \sim$\ }}}
\def\gaeq{\lower.5ex\hbox{\fiverm{\ $\buildrel > \over \sim$\ }}}
\def\boxit#1{\vbox{\hrule\hbox{\vrule\kern3pt
     \vbox{\kern3pt#1\kern3pt}\kern3pt\vrule}\hrule}}
%
%

%
%

%
%

\def\leftdisplay#1$${\leftline{\noindent$\displaystyle{#1}$}$$}
%
%
%
%
 \font\medbx=cmbx10 scaled \magstep2

 \font\twelvebf=cmbx12 
 \font\twelvett=cmtt12 at 12pt
 \font\twelverm=cmr12 at 12pt    \font\ninerm=cmr9
\font\sevenrm=cmr7    \font\fiverm=cmr5
 \font\twelvei=cmmi12 at 12pt    \font\ninei=cmmi9
\font\seveni=cmmi7    \font\fivei=cmmi5
 \font\twelvesy=cmsy10 at 12pt   \font\ninesy=cmsy9
\font\sevensy=cmsy7   \font\fivesy=cmsy5
 \font\twelveit=cmti12 at 12pt
 \font\twelvesl=cmsl12 at 12pt
 \font\tenex=cmex10
 \font\teni=cmmi10    \font\tensy=cmsy10
 \font\tentt=cmtt10   \font\tenit=cmti10
 \font\tensl=cmsl10
%
%
\def\BJBtwelvepoint{\def\rm{\fam0\twelverm}
 \textfont0=\twelverm \scriptfont0=\ninerm \scriptscriptfont0=\sevenrm
 \textfont1=\twelvei \scriptfont1=\ninei \scriptscriptfont1=\seveni
 \textfont2=\twelvesy \scriptfont2=\ninesy \scriptscriptfont2=\sevensy
 \textfont3=\tenex \scriptfont3=\tenex \scriptscriptfont3=\tenex
 \def\it{\fam\itfam\twelveit}%
 \textfont\itfam=\twelveit
 \def\sl{\fam\slfam\twelvesl}%
 \textfont\slfam=\twelvesl 
 \def\bf{\fam\bffam\twelvebf}%
 \textfont\bffam=\twelvebf 
 \def\tt{\fam\ttfam\twelvett} 
 \textfont\ttfam=\twelvett
 \font\bigit=cmti12 scaled \magstep2
 \baselineskip 14pt%
 \abovedisplayskip 14pt plus 3pt minus 10pt%
 \belowdisplayskip 14pt plus 3pt minus 10pt%
 \abovedisplayshortskip 0pt plus 3pt%
 \belowdisplayshortskip 8pt plus 3pt minus 5pt%
 \parskip 3pt plus 1.5pt
 \setbox\strutbox=\hbox{\vrule height10pt depth4pt width0pt}%
 \rm}
%
%
\def\BJBtenpoint{\def\rm{\fam0\tenrm}
 \textfont0=\tenrm \scriptfont0=\sevenrm \scriptscriptfont0=\fiverm
 \textfont1=\teni \scriptfont1=\seveni \scriptscriptfont1=\fivei
 \textfont2=\tensy \scriptfont2=\sevensy \scriptscriptfont2=\fivesy
 \textfont3=\tenex \scriptfont3=\tenex \scriptscriptfont3=\tenex
 \def\sl{\fam\itfam\tenit}%
 \textfont\itfam=\tenit
 \def\sl{\fam\slfam\tensl}%
 \textfont\slfam=\tensl 
 \def\bf{\fam\bffam\tenbf}%
 \textfont\bffam=\tenbf 
 \def\tt{\fam\ttfam\tentt}%
 \textfont\ttfam=\tentt
 \baselineskip 12pt%
 \abovedisplayskip 14pt plus 3pt minus 10pt%
 \belowdisplayskip 14pt plus 3pt minus 10pt%
 \abovedisplayshortskip 0pt plus 3pt%
 \belowdisplayshortskip 8pt plus 3pt minus 5pt%
 \parskip 3pt plus 1.5pt
 \setbox\strutbox=\hbox{\vrule height8.5pt depth3.5pt width0pt}%
 \rm}

\BJBa4
\BJBtenpoint
\noindent
\vglue 2.0cm
\centerline{\medbx The Clustering of K $\sim$ 20 Galaxies}
\centerline{\medbx on 17 Radio Galaxy Fields} 
\vskip 1.5cm
\centerline{Nathan Roche, Steve Eales}
\vskip 0.4cm 
\centerline{Department of Physics and Astronomy,
University of Wales Cardiff,}

\centerline{P.O. Box 913, Cardiff CF2 3YB, Wales.}
\vskip 1.0cm
\centerline{Hans Hippelein}
\vskip0.4cm
\centerline{Max Planck Institut f\"ur Astronomie, K\"onigstuhl 17,
69117 Heidelberg, Germany.}
\vskip 1.5cm
\centerline{\bf Abstract}
We investigate the angular correlation function, $\omega(\theta)$, of the galaxies detected in the $2.1\mu \rm m$ $K^{\prime}$-band on 17 fields (101.5 $\rm arcmin^2$ in total), each containing a $z\sim 1.1$ radio galaxy. There is a significant detection of galaxy clustering at limits of $K\sim 20$, with a
$\omega(\theta)$  amplitude 
similar to that estimated by Carlberg et al. (1997) at $K=21.5$. The
$\omega(\theta)$ amplitudes of these $K$-limited samples are
higher than expected 
from simple models which fit the faint galaxy clustering in the
blue and red passbands, but consistent  with a pure luminosity evolution model if clustering is stable ($\epsilon =0$) and the correlation function of early-type galaxies is steeper than that of spirals.

We do not detect a significant cross-correlation between the radio galaxies and the other galaxies on these fields. The upper limits on the cross-correlation are consistent with a mean clustering environment of Abell class 0 for $z\sim 1.1$
radio galaxies, similar to that observed for radio galaxies at $z\sim 0.5$, but would argue against an Abell class 1 or richer environment. As 
Abell 0 clustering around the radio galaxies would not significantly increase the $\omega(\theta)$ amplitude of galaxies on these fields, stable clustering with a steep $\xi(r)$ for E/S0 galaxies appears to remain the most
likely interpretation of the $\omega(\theta)$ amplitude.

At $K\leq 20$, the number of galaxy-galaxy pairs of 
2--3 arcsec separations exceeds the random expectation by a factor 
of $2.15\pm 0.26$.
This excess of close pairs is comparable to that previously 
reported on $R$-band data, and suggests at least a tripling of the local merger rate at $z\sim 1$. 
\bigskip
\noindent {\bf 1. Introduction}
\smallskip
The clustering of galaxies in 3D space can be described 
by their
two-point correlation function, $\xi(r)$, which
approximately follows the power-law
$$\xi(r,z)=({r\over r_0})^{-\gamma}(1+z)^{-(3+\epsilon)}$$
where $r_0$ is the correlation radius, $\gamma$ the slope, and $\epsilon$ 
parameterizes the evolution of clustering.
Observations 
suggest $\gamma\simeq 1.8$ and for nearby field galaxies,
Loveday et al. (1995) estimate
$r_0= 4.4$ $h^{-1}\rm Mpc$ for
spirals and irregulars ($h$ is $H_0$ in units of 100 km$\rm s^{-1}Mpc^{-1}$)
and $r_0=5.9$ $h^{-1}\rm Mpc$ for
E/S0 galaxies.
 There is still much uncertainty about the rate of clustering evolution, $\epsilon$. Stable clustering corresponds to $\epsilon=0$, while comoving clustering
is $\epsilon=-(3-\gamma)=-1.2$ for $\gamma=1.8$. Clustering which collapses at the rate at which the Universe expands is then $\epsilon=1.2$. Models
(e.g. Colin, Carlberg and Couchman 1997) and direct observations 
(e.g. Le Fevre et al. 1996; Shepherd et al. 1997; Carlberg et al. 1997)
generally span the $0\leq \epsilon\leq 1.5$ range.

The clustering of galaxies on the sky is described by the angular correlation function $\omega(\theta)$, which is a power-law of slope $\theta^{-(\gamma-1)}$ and amplitude given by the Limber's formula integration of $\xi(r,z)$ over the galaxy redshift distribution $N(z)$. The $\omega(\theta)$ amplitude is therefore
sensitive to $r_0$, $\epsilon$ and $N(z)$.
At magnitude limits of $B\geq 23.5$, the $\omega(\theta)$ amplitude was found to fall some $\sim 0.5$ dex  below the predictions of models with $\epsilon=0$ and no luminosity evolution for the galaxies (e.g. Brainerd, Smail and Mould 1995; Roche et al. 1996a and references therein). This was interpreted as evidence that $N(z)$ at these magnitudes is
more extended than the non-evolving prediction, indicating an  
evolutionary brightening of $L\sim L^*$ galaxies at higher redshifts, but  alternatively could mean that the intrinsic clustering of galaxies was much weaker at early epochs (i.e. $\epsilon\geq 1$).
Spectroscopic observations (e.g. Cowie et al. 1995, 1996; Steidel et al. 1996) have now found many galaxies at $z\sim 1$--3 and confirmed there is a brightening of the galaxy $L^*$ at these redshifts, so the aim of studying faint galaxy clustering may now be changing from its use as an indirect probe of $N(z)$ towards the investigation of clustering evolution and the relation between galaxy evolution and clustering environment.

 Recently it has become possible to observe sufficiently large areas in
 the $2.2\mu \rm m$ $K$-band to measure the $\omega(\theta)$ 
amplitude for
 $K$-limited galaxy samples. This may give different results from 
$\omega(\theta)$ analyses of galaxies selected at shorter wavelengths, as (i)
$K$-limited surveys will contain a much higher proportion of early-type galaxies, and (ii) $K$-band observations are less
sensitive to evolution of the star-formation rate, and so may be more useful
in constraining $\epsilon$ rather than the rate of $L^*$ evolution. 
Baugh et al. (1996) measured $\omega(\theta)$ for galaxies on two
large fields (totalling 9 $\rm deg^2$) at limits of $K=15$ and $K=16$, and Carlberg et al. (1997) obtained a $\omega(\theta)$ amplitude from a 27 $\rm arcmin^2$ area at a much deeper limit of $K=21.5$.
 As part of a program to investigate radio galaxy evolution (Eales et al. 1997).
we have obtained $K$-band images of areas centred on 17 high redshift radio galaxies. These images cover a total
of 101.5 $\rm arcmin^2$, with galaxies reliably detected to $K\sim 20$, and may provide an estimate of the $\omega(\theta)$ amplitude at this intermediate limit. However, as some high redshift radio galaxies lie within rich
galaxy clusters (e.g. Hill and Lilly 1991), we must consider the possible
effect of the radio galaxies on $\omega(\theta)$.

We assume $H_0=50$ km $\rm s^{-1} Mpc^{-1}$ and $q_0=0.05$ throughout.
In Section 2 of this paper we describe the observational data, and in Section 
3 the detection and number counts of the galaxies. Section 4 describes the measurement of the galaxy $\omega(\theta)$ and compares with models
and previous observations. In Section 5 we investigate
the clustering environment of the radio galaxies by cross-correlating
 with the surrounding galaxies, and also discuss the effects of the radio galaxies
on $\omega(\theta)$. In Section 6 we investigate the numbers of close pairs of
galaxies on these images, using the methods of Woods et al. (1995), which may
provide information on merging. In Section 7 we discuss the implications of these results and summarize the main conclusions in Section 8.    
\bigskip
\noindent {\bf 2. Observational Data}
\smallskip
 The observational dataset of this paper consists of images of 17 fields, each centred on a high redshift radio galaxy. The radio galaxies were chosen
from the 6C survey (see e.g. Eales 1985), except for 5C6.8 which is a radio-loud
QSO in the same redshift range.
Spectroscopic redshifts were obtained by Rawlings et al. (1997) for all but
two of the radio sources, giving $z_{mean}=1.12$ for the observed 15.

All these
observations were made in the $2.1\mu \rm m$ $K^{\prime}$-band (Wainscoat and Cowie 1992) rather than the standard  $2.2\mu \rm m$ $K$-band.
The first 12 fields were observed over the period 18--20 January 1995 using
the Redeye camera on the Canada-France Hawaii Telescope (CFHT),
a $256\times 256$ HgCdTe mosaic with a pixelsize of 0.5 arcsec. Each field was imaged nine times, with the camera being
offset by 8 arcsec between exposures in a $3\times3$ grid, and total 
exposure times from 26 to 47 minutes per field. 
Seeing is estimated as $FWHM\simeq 0.9$--1.2
arcsec from stars on the reduced images.
 These images were previously used in studying the luminosity 
evolution (Eales et al. 1997), morphology and surface brightness evolution (Roche et al. 1997b) of
the radio galaxies.
The second part of the data consists of a further 5 radio galaxy fields, 
observed using the Magic camera on the 3.5m Calar Alto Telescope, which contains a similar detector to that in the Redeye camera but here gives 
a larger pixelsize of $0.81^{\prime\prime}$. These five fields were 
observed in February 1997, for total exposure times of 40 to 66 minutes per field. The resolution  
is estimated as $FWHM\simeq 1.6$--1.9 arcsec, i.e. a little 
poorer than for the Redeye data.

Table 1 lists the positions of the radio galaxies, with their names, redshifts and $K$ magnitudes 
(as measured from this data).  As our analysis will require at least approximate redshifts for all the radio galaxies, for the two objects not yet observed spectroscopically we estimate very approximate redshifts from their $K$ magnitudes and the best-fitting model for the 6C galaxy $K-z$ relation (Eales et al. 1997).
These estimated redshifts and any quantities derived from them are shown on the tables as
bracketed numbers.

\bigskip
\noindent {\bf 3. Galaxy Detection and Number Counts}
\smallskip
The Redeye images were initially processed using the IRAF package, while the Magic images were separately processed using MIDAS. The stacked and 
flat-fielded  images were analysed further using the
  Starlink KAPPA and PISA packages.
 On the Redeye images, the shifting of the camera between exposures produced regions at the edges with noticeably poorer signal-to-noise than the
central areas, so we used only the
central $224\times 224$ pixel areas. 
The total area of the 12 Redeye fields is then 41.8 $\rm arcmin^2$.
On the Magic images, we used the full $256\times 256$ pixel areas, giving
a total area for the 5 fields of 59.7 $\rm arcmin^2$. The full
dataset of this paper covers 101.5 $\rm arcmin^2$. Note that 
for the assumed cosmology, 224 Redeye pixels corresponds
at $z=1$ to 1.18 Mpc, and 256 Magic pixels to
2.19 Mpc.

   The Starlink PISA (Position, Intensity and Shape Analysis) package, developed by M. Irwin, P. Draper and N. Eaton, was used to detect and catalog the
objects on each field. Our chosen 
detection criterion was that an object must exceed an intensity threshold of
$1.5\sigma$ above the background noise ($\sigma$ being separately determined by
PISA for each of the frames) in at least 6 connected
pixels on the Redeye inages and 4 connected pixels on the Magic images. Table 1 gives these detection thresholds in terms of $K$ mag $\rm arcsec^{-2}$. PISA
was run with the deblend option, so that objects connected at the detection threshold may be split if they
are separate at a higher threshold.

 On the 12 Redeye images, PISA detected a total of 1053 objects, although
 43 of these appeared to be spurious  detections caused by streaks on the
images, bright stars etc., and
 were excluded from the catalog. There were a total of 1039 
detections on the Magic images, of which 11 were excluded as obviously spurious.
To estimate fluxes for the detected objects, 
we used the `total magnitudes' option within PISA, which counts the
photons above the background level in elliptical apertures fitted to the intensity profile of each 
individual detection.

 Observations of $K$-band standard stars provided photometric zero-points 
of $K=21.28$ for 1 count $\rm sec^{-1}$ on the Redeye data and
$K=21.90$ for 1 count $\rm sec^{-1}$ on the Magic data, but due to the difference between the $K$ and $K^{\prime}$ passbands,
 these will only be exactly correct for objects with the same
$K^{\prime}-K$ as the standard stars. Radio galaxies 
at $1\leq z\leq 1.4$ will be redder than the 
calibration stars by an estimated
$\Delta(K^{\prime}-K)\simeq 0.13$ mag (Eales et al. 1997), as will  
early-type galaxies at similar redshifts (which make up a large part of the sample) so these zero-points were adjusted by $-0.13$ mag for all detections.
Total $K$ magnitudes were estimated for all detected objects. The radiogalaxies were identified on each image, and Table 1 lists 
their $K$ magnitudes as measured from this 
data.
Star-galaxy separation was only possible to $K\sim 17$-18.
On the 17 fields, a total of 71 objects 
were classed as stars, by eye and using size parameters from PISA, and fainter detections were assumed to be galaxies.

Figure 1 shows the galaxy number counts derived from these 17 fields, in 
$\Delta(K)=0.5$ mag bins, compared with some previously published $K$-band galaxy counts. At $15<K<20$ our number counts are consistent with  previous surveys, perhaps a little higher than average. Beyond $K\sim 19.5$--20,
 the counts level out and begin to fall off as the
detection becomes increasingly incomplete. 
To $K=20$, a total of 1357 galaxies are detected, giving a surface density of 
48130 $\rm deg^{-2}$. This is
within the range of previous observations at this magnitude limit, being
higher than the 33710 $\rm  deg^{-2}$ in the Hawaii Deep Survey (Gardner,
Cowie and Wainscoat 1993), but slightly lower than the 
52000 $\rm deg^{-2}$ in the smaller but deeper survey of Moustakas et al. (1997).
 The counts are also compared with two models, based on  the spectral evolution models of Charlot and Bruzual (1993) and computed
in the $K$-band. The non-evolving model is that described in Roche et al. (1996b); the pure luminosity evolution (PLE) model has the same zero-redshift luminosity functions but with luminosity evolution derived from the `SLE model' of Roche et al. (1997a).  
The counts show some indication of an excess over the non-evolving model but there is 
much scatter between the different observations, so that both models may be 
consistent with current data.  
\bigskip
\noindent{\bf 4. The Galaxy Angular Correlation Function $\omega(\theta)$}
\smallskip
\noindent {\bf 4.1 Calculating $\omega({\theta})$}
\smallskip
The galaxy-galaxy angular
correlation function $\omega_{gg}(\theta)$ was calculated following the  methods of Roche et al. (1996a). 
Each of the 17 fields was analysed separately, with no distinction made between  the radio galaxies and the other galaxies. If $N_g$ is the number of galaxies on
a field brighter than a chosen magnitude limit, there will be ${1\over2}N_g(N_g-1)$ possible 
galaxy-galaxy pairs. The separations of all these pairs are counted in bins of width $\Delta ($\rm log
$\theta)=0.2$. We placed 10000 points randomly over the field area and 
similarly counted the galaxy-random pairs, taking the real galaxies as the centres. The angular correlation function $\omega_{gg}(\theta)$ is then calculated as
$$\omega_{gg}(\theta_i)={N_{gg}(\theta_i)\over N_{gr}(\theta_i)}
{2N_r\over(N_g-1)} - 1 $$ where $N_{gg}(\theta_i)=$ number of
galaxy-galaxy pairs in bin i, $N_{gr}(\theta_i)=$ number of
galaxy-random pairs in bin i and $N_r=$ number of random points (10000
here).
The integral constraint, an effect caused by the finite field sizes, causes the observed $\omega(\theta)$ to be 
negatively offset from the true correlation function $\omega_T({\theta})$,
by a constant equal to $\omega_T({\theta})$ doubly integrated over the
field area $\Omega$,
$$C=\int\int \omega_T(\theta) d\Omega_1 d\Omega_2$$
Assuming $\omega_T(\theta)=A\theta^{-0.8}$, where $A$ is the power-law amplitude at one degree, $C=36.0A$ for the Redeye fields and $C=22.0A$ for the Magic 
fields, so that the observed $\omega(\theta)$ corresponds to
$A(\theta^{-0.8}-36.0)$ and $A(\theta^{-0.8}-22.0)$ respectively.

The individual field $\omega(\theta)$ were averaged to give mean 
$\omega(\theta)$ for the
Redeye and Magic data. These are shown on Figure 2, with field-to-field errors, at magnitude limits of
$K=19.5$, 20.0 and 20.5 (the result for the deepest limit may be less reliable due to incompleteness).

 The $\omega(\theta)$ from the Redeye data shows a positive signal down to
 smaller separations, $2.0<\theta<3.2$ arcsec, than that from the Magic data 
which is only positive down to $3.2<\theta<5.0$ arcsec. This is presumably due to the difference in resolution.
The $\omega(\theta)$ amplitude of the Redeye data was estimated by fitting
a function `$A(\theta^{-0.8}-36.0)$' over the $2.0\leq \theta \leq 126$
arcsec range of separations, using least-squares fitting weighted with the
field-to-field errors. The error on $A$ was estimated from the field-to-field
scatter in the amplitudes derived by fitting to $\omega(\theta)$ of the individual fields. 
The correlation function of the Magic data galaxies falls below zero at larger separations, suggesting that it is affected by sensitivity gradients, so
 we estimate $A$ by fitting `$A(\theta^{-0.8}-22.0)$' over just the
$3.2\leq \theta \leq 8.0$ arcsec range. 

Table 2 gives the correlation function amplitudes estimated from the two datasets, with an error-weighted combined value.
When all our data is combined, galaxy clustering is detected with at least
$\sim 4\sigma$ significance at these magnitude limits.
\smallskip
\noindent {\bf 4.2 Interpretation of $\omega(\theta)$}
\smallskip
Figure 3 shows the combined $\omega(\theta$) amplitudes from this data and 
those from other $K$-limited galaxy surveys,
as a function of  $K$ limit.
Baugh et al. (1996) investigated the galaxy $\omega(\theta)$ at relatively shallow limits
of $K=15$ and $K=16$ on two (NEP and NGP) fields totalling 9 $\rm deg^2$, where
it appeared consistent with a $\theta^{-0.8}$ power-law out to
$\theta\sim 1$ degree, and its amplitude was estimated by fitting this power-law at $18<\theta<216$ arcsec. Carlberg et al. (1997) measured the galaxy clustering
at a much deeper limit of $K=21.5$ on a much smaller
area of 27 $\rm arcmin^2$, over angular scales of
approximately $1.5<\theta<150$ arcsec.
 We plot their result without their rather
large correction for star contamination, as this may 
be an overestimate (being based on the
proportion of stars at brighter magnitudes), and as no such correction has been applied to our results. 

Our $\omega(\theta)$ amplitudes are much lower than those at $K=15$--16 but consistent with the $K=21.5$ result. We compare these results with four models; 

A: Our non-evolving model with $\epsilon=0$,  
$r_0=5.9$ $h^{-1}$ Mpc for E/S0 galaxies and  $r_0=4.4$ $h^{-1}$ Mpc for later types
(from Loveday et al. 1995), 
the geometric mean $r_0=5.1$ $h^{-1}$ Mpc for the cross-correlation between the two (see Roche et al. 1996a), and an additional weighting term giving more luminous ($M_B<-20.5$) galaxies
twice the strength of clustering as galaxies of lower luminosity (as
estimated by Loveday et al. 1995).
 
B: Our PLE model with $\epsilon=0$. This predicts lower $\omega(\theta)$
amplitudes at fainter limits due to the more extended galaxy $N(z)$.
 Figure 4 shows the $N(z)$ predicted by the PLE model for $K\leq 20$. This
has a mean redshift $z_{mean}=1.09$, similar to that of the radio galaxies, and  a $47.5\%$ fraction of E/S0 galaxies, which is much higher than in a $B$-limited sample of similar depth
(e.g. Figure 15 of Roche et al. 1996a). This model is successful in fitting the 
$\omega(\theta)$ scaling in the blue and red passbands (Roche et al. 1996a).   

C: The PLE model with clustering becoming weaker in the past, at a rate
$\epsilon=1.2$, a rate of clustering evolution towards the maximum of the observed and predicted range (see
Introduction).

D: The PLE model with $\epsilon=0$ and a higher normalization and and steeper slope for the $\xi(r)$ of early-type galaxies. Guzzo et al. (1997) estimate
 $r_0=8.35\pm 0.75$ $h^{-1}$ Mpc and $\gamma=2.05\pm 0.09$ for
 the clustering of giant ($M_B<-21.5$ approx.) E/S0 galaxies. 
Model D adopts this higher estimate, with the same spiral galaxy clustering
as the other models, dwarfs half as clustered as giant galaxies and the
geometric mean $r_0=5.8 h^{-1}$ Mpc and $\gamma=1.925$ for the E/S0-spiral cross-correlation. Relative to model B, the stronger E/S0 clustering  makes 
little difference ($<10\%$) to $\omega(\theta)$ at $\theta \sim 1$ degree, but greatly increases $\omega(\theta)$ at very small separations.  We plot the amplitude which would be
measured at $\theta=5$ arcsec, if a $\theta^{-0.8}$ power-law 
is assumed. This should
 be approximately valid for comparison with our results and those
of Carlberg et al. (1997), but should not be compared with the Baugh et al. (1996) results which were measured at larger separations. 

The non-evolving model A gives a $\omega(\theta)$ amplitude consistent with our result but may slightly overpredict the
amplitude at $K=16.0$ and underpredict it at $K=21.5$. However, there is
evidence for luminosity evolution from spectroscopy (Cowie et al. 1996), which
confirms that many $19<K<20$ galaxies are at $1<z<2$, as in our PLE models.
Considering the evolving models, model B 
fits the $\omega(\theta)$ amplitude at $K=16$ but underpredicts at $K\geq 19.5$. At $K=20$, our measured amplitude of 
$13.33\pm 2.32\times 10^{-4}$  exceeds the model B prediction of
$5.7\times 10^{-4}$ by $\sim 3\sigma$, so the discrepancy is significant, especially when the Carlberg et al. (1997) result is also considered.
Model C greatly underpredicts the observed amplitudes, 
arguing against any model in which clustering is much weaker in the past  (i.e. $\epsilon>1$). In model D,
the stronger E/S0 clustering increases the $\omega(\theta)$
amplitude at  $\theta\sim 5$ arcsec by $\sim 0.4$ dex, so that it is at least as high as our observed values. 

These results suggest that the $\omega(\theta)$ scaling in the $K$-band is consistent with a $L^*$ evolution model (which also fits the galaxy counts)
 if (i) clustering is approximately stable and (ii) the $\gamma$ and 
$r_0$ of the E/S0 galaxy $\xi(r)$ are closer to the high 
Guzzo et al. (1997) estimates. However, we must also consider, in Section 5 below, the alternative possibility that  clusters around the radio galaxies on these
fields produce an enhancement of the galaxy $\omega(\theta)$.
\bigskip
\noindent {\bf 5. Clustering Associated with the Radio Galaxies}
\smallskip
\noindent {\bf 5.1 Introduction}
\smallskip
It is possible that some of the detected galaxies belong to 
rich clusters centred on the radio galaxies. Within rich clusters, the
amplitude of $\xi(r)$ may
be much higher than for field galaxies.
The strength of
clustering at 1 Mpc separations can also be parameterized as $B_{gg}$, equivalent to
$r_0^{\gamma}$.
For $H_0=50$ km $\rm s^{-1}Mpc^{-1}$ and $\gamma=1.8$, the Loveday et al. (1995)
estimates of the field galaxy $r_0$
 correspond to $B_{gg}=85.0$ for E/S0s and $B_{gg}=50.1$ for spirals, whereas the
average $B_{gg}$ is estimated as 289, 693 and 990 for Abell class
0, 1 and 2 clusters respectively (Prestage and Peacock 1988, 1989).
We consider the galaxy clustering around the radio galaxies in terms of a 3D radiogalaxy-galaxy cross-correlation function,
assumed to have the same form as $\xi(r)$, with an amplitude $B_{rg}$. If the radio galaxies lie within rich clusters, $B_{rg}$ will be similar to $B_{gg}$ in the cluster cores. 

At low redshifts, Longair and Seldner (1979) found 
a mean $B_{rg}$ of only $48\pm 54$ for FRII emission line 
radio galaxies, while Prestage and Peacock (1988, 1989) found a mean
$B_{rg}$ of $132\pm 43$, indicating an environment similar to that of randomly chosen field galaxies.
The radio galaxies on these 17 fields are all
 luminous FRII sources, so if their environment is similar to that of their low-redshift counterparts, they would randomly sample the galaxy distribution  and their presence would have no effect on the galaxy $\omega(\theta)$.

However, at higher redshifts many FRII galaxies
do lie in rich clusters.
Yates, Miller and Peacock (1989)
 found that a mean $B_{gr}=101\pm 43$ for powerful radio galaxies at $0.15<z<0.35$ increased to $B_{gr}=252\pm 81$ for those at $0.35<z<0.82$, and 
Hill and Lilly (1991) confirmed this with a larger sample, finding  
a mean richness  corresponding to $B_{gg}=291\pm 45$ (Abell class 0) for 
high-luminosity radiogalaxies at $0.35<z<0.55$.
Yee and Green (1987) found that the mean cross-correlation of radio-loud QSOs with surrounding galaxies increased from $B_{qg}=161\pm 80$ for 10 QSOs at $0.3<z<0.5$ to $B_{qg}=543\pm 142$ (Abell class 0-1) for 9 at $0.55<z<0.65$.  
 Yee and Ellingson (1993) suggested that these observations can
be explained if the luminosity of radio sources in clusters evolves much more rapidly 
($\tau\simeq 1$ Gyr) out to $z\sim 0.6$ than that of their field counterparts.

If the $z\sim 1.1$ radio galaxies on these fields lie in clusters, we might expect to see both an enhanced 
$\omega(\theta)$ amplitude for the galaxies, and a clustering of galaxies near the radio galaxy positions.
However, both these effects will be greatly diluted by
 the presence of non-cluster galaxies, as we consider below.  
\smallskip
\noindent {\bf 5.2 Calculating $\omega_{rg}(\theta)$}
\smallskip
We calculate a radiogalaxy-galaxy cross-correlation function $\omega_{rg}(\theta)$ for each of the 17 fields, by summing the number of
$K\leq 20$ galaxies and the number of random points in annular bins of width $\Delta(\rm log~\theta)=0.2$ centred on the radiogalaxy. The cross-correlation 
function is then calculated as
 $$\omega_{rg}(\theta)={N_{gg}(\theta_i)\over N_{gr}(\theta_i)}
{N_r\over N_g} - 1 $$
 where $N_{gg}(\theta_i)=$ number of
radiogalaxy-galaxy pairs in bin i, $N_{gr}(\theta_i)=$ number of
radiogalaxy-random pairs in bin i and $N_r=$ number of random points (10000
here).

 Figure 5 shows a mean
cross-correlation function, with field-to-field errors, obtained by averaging
$\omega_{rg}(\theta)$ for all 17 fields, with equal weighting for each field.  This will be negatively offset by an integral constraint, which
for the assumed power-law
$\omega_{rg}(\theta)=A_{rg}\theta^{-0.8}$ can be estimated by the single integration over the field
area $$C=\int A_{rg}\theta^{-0.8} d\Omega$$
giving $C=40.8A$ for the Redeye fields and $C=25.3A$ for the Magic 
fields. For the mean $\omega_{rg}(\theta)$ of the 17 fields, we assumed a mean integral constraint $C\simeq 36.2A$, and fit $A_{rg}(\theta^{-0.8}-36.2)$ over the $3.2<\theta<79.6$ arcsec range of separations. An error is estimated by performing separate fits for each individual field (with the appropriate integral constraints) and finding the scatter in the individual field amplitudes.

 At $K\leq 20.0$, we estimate
 $A_{rg}=9.21\pm 15.82\times 10^{-4}$ for the 17 radio galaxies. There is no detection of a significant signal.
\smallskip
\noindent {\bf 5.3 Interpretation}
\smallskip
This result is interpreted by modelling the expected cross-correlation as
a function of $B_{rg}$, using methods
similar to Hill and Lilly (1991). Assuming a radiogalaxy-galaxy cross-correlation with $\gamma=1.8$ and $\epsilon=0$, we  calculate

\noindent (i) The ratio of the amplitude, $A_{cl}$, of the angular correlation function of
the cluster galaxies at one degree to the amplitude 
of their two-point correlation function at 1 Mpc, $B_{rg}$.
 Limber's formula integration,
 evaluated at the cluster redshift only gives
$$A_{cl}=3.679({\pi\over 180})^{-0.8}d_{A}(z)^{-0.8}{dr\over dz}(z)B_{gr}(1+z)^{-3}$$
where $d_{A}(z)$ is the angular diameter distance at the radio galaxy redshift
and ${dr\over dz}$ the derivative of proper distance with respect to redshift.
This gives ${A_{cl}\over B_{rg}}=1.74\times 10^{-5}$ at $z=1$, decreasing slowly with redshift.

\noindent (ii) $\phi(K\leq 20)$, the number density per unit volume
of galaxies which would be brighter than $K=20$ in apparent magnitude, in our
PLE model, at the radio galaxy redshift, for a mean field
density of galaxies. Our model gives $\phi(K\leq 20)\simeq 2.1\times 10^{-3}
\rm Mpc^{-3}$ at $z=1$, decreasing steeply with redshift, e.g. by
a factor of $\sim 4$ to $z=1.8$.

\noindent (iii)  The integral, $I_{rg}$, of the two-point radiogalaxy-galaxy cross-correlation function, $\xi_{rg}(r,z)$, over the field area (centred on the radio galaxy) and
along the line of sight to the observer. This is
$$I_{rg}=B_{rg}\int\int r^{-1.8}(1+z)^{-3} d\Omega dz$$
where $d\Omega$ represents the integration over the field area, centred on the
radio galaxy, and $dz$ the integration over redshift, in both cases using
 units of Mpc.
We estimate  $I_{rg}\simeq 11B_{rg}$ for the Redeye fields and 
$I_{rg}\simeq 22B_{rg}$ for the Magic fields, with only a weak dependence on redshift.

The mean number of $K\leq 20$ 
galaxies in each radio galaxy cluster, $N_{cl}$, can be then
estimated as 
$N_{cl}=1+ \phi(K\leq 20)I_{rg}$
 where the addition of 1 accounts for the radio galaxy itself.

The amplitude of the angular cross-correlation between the radio galaxies and
the surrounding galaxies on the images, $A_{rg}$, will be lower than $A_{cl}$
due to dilution by non-cluster (foreground and background) galaxies. 
 If the total
number of $K\leq 20$ galaxies detected on a field is $N_{T}$, $A_{rg}={N_{cl}\over N_{T}}A_{cl}$.
For this data we estimate 
${N_{cl}\over N_{T}}\simeq 4.0\times 10^{-4}B_{rg}$, only $\sim 0.13$ if the mean $B_{rg}$ is similar to the estimates at $z\sim 0.5$.
Both $N_{cl}$ and $A_{cl}$ depend on $B_{rg}$, so the observed cross-correlation becomes,
$$A_{rg}={[1+ \phi(K\leq 20)I_{rg}]\over N_{T}}({A_{cl}\over B_{rg}})B_{rg}$$
Averaging our estimates of these quantities for all 17 fields,
$$A_{rg}\simeq 3.00\times 10^{-7}B_{rg}+6.75\times 10^{-9}B_{rg}^2$$   

which is plotted on Figure 6a. Our best-fitting $A_{rg}$ 
corresponds to $B_{rg}=348$, consistent with the same  mean clustering
environment as estimated  (Hill and Lilly 1991) 
for radio galaxies at $0.35<z<0.55$. As there was no significant signal
in the cross-correlation, our results give no lower limit on $B_{rg}$, but 
do give estimated $1\sigma$ and $2\sigma$ upper limits of $B_{rg}=587$
and  $B_{rg}= 756$, so appear to  disfavour a richer
mean clustering environment of Abell class 1--2.

The effect of clustering around radio galaxies on the galaxy
$\omega({\theta})$ will be subject a more
complex dilution, proportional to the square of the fraction
of non-cluster galaxies. If the $\omega(\theta)$ amplitude of the field galaxies is $A_{f}$ and we neglect any cross-correlation between the field and cluster galaxies (reasonable as only a very small proportion of the field
galaxy $N(z)$ will overlap with the cluster), $$A=({N_{cl}\over N_{T}})^2A_{cl}+(1-{N_{cl}\over N_{T}})^2A_{f}$$
Figure 6b shows a model for the mean $A$ on these fields as a function of $B_{gr}$, assuming
$A_{f}=5.7\times 10^{-4}$ at $K\leq 20$ from the $\epsilon=0$ PLE model. This
model suggest the  $\omega(\theta)$ amplitude will be
insensitive to Abell class 0 clustering around the radio galaxies, but stronger clustering of $B_{rg}>400$ will begin to make an impact. The difference between
 the observed $\omega(\theta)$ at $K\leq 20$ and the PLE prediction could be accounted for a mean $B_{rg}=664^{+53}_{-62}$, although this is only marginally
consistent with the upper limits on the radiogalaxy-galaxy cross-correlation.
The $\omega(\theta)$ amplitude would in itself set a $2\sigma$ upper limit
of $B_{rg}=762$.

In summary, the observations are consistent with a mean clustering environment of Abell class 0 for 6C radiogalaxies at $z\sim 1.1$, but may argue against Abell class 1 or richer clustering. At this point, we cannot exclude the possibility that  the high galaxy $\omega(\theta)$ amplitude relative to our model B is the result of strong clustering around the radio galaxies, but this hypothesis is not favoured. We discuss this further in Section 7.
\bigskip
\noindent {\bf 6. Counting Close Pairs of Galaxies}
\smallskip
\noindent {\bf 6.1 Introduction}
\smallskip
We now consider the numbers of very close pairs of galaxies on these images, i.e. pairs which are likely to be physically interacting.  
Yee and Ellingson (1995), investigating the 3D clustering of spectroscopically
observed $r<21.5$ ($z_{mean}=0.38$)
field galaxies,  estimated that $8.5\%$  belonged to interacting  pairs of physical separation $r<20$ $h^{-1}$ kpc. Similarly, Patton et al. (1997) estimated that $7.1\pm 1.4\%$ of the galaxies in a larger $z_{mean}=0.33$ survey  were in physical pairs 
of separation $r<20$ $h^{-1}$ kpc, compared to $4.3\pm 0.3\%$ locally.

This excess of close galaxy pairs has also been seen in the $\omega(\theta)$ of 
non-spectroscopic surveys. 
Infante et al. (1996) found that the $\omega(\theta)$ of $19<R<21.5$
galaxies ($z_{mean}\simeq 0.35$) followed a double $\theta^{-0.8}$  power-law,
with a higher amplitude at $\theta<6$ arcsec ($r<19$ $\rm h^{-1} kpc$ at the mean redshift) than at larger scales.
The number of galaxy pairs at $2<\theta<6$ separation exceeded the random expectation by a factor of 1.637,
and exceeded the number expected from an 
inward extrapolation of the larger scale $\omega(\theta)$ by 1.361. 
 It was estimated from this that 
$10.04\pm 0.17\%$ of the galaxies were in  $2<\theta<6$ arcsec pairs, after subtracting the random expectation. Subtracting the
extrapolated $\omega(\theta)$ prediction would reduce this to $6.4\%$, which might better represent the fraction of merging galaxies.

The $\omega(\theta)$ from our much smaller survey (Figure 2) might suggest an excess over the fitted $\theta^{-0.8}$ power-law at $2<\theta<3.2$ arcsec,
for the Redeye data only, although only of 1--$2\sigma$ significance. However,
we can apply a statistically more powerful test for an excess of
close pairs, which makes optimal use of the magnitude information.
Woods et al. (1995) describe a method in which a probability $P$ of occurring by chance (in a random distribution
of galaxies) is calculated for each  galaxy-galaxy pair. The probability is estimated as
$$P=\int^{\theta}_{\beta}{\rm exp}(-\pi\rho\alpha^2) d\alpha$$
where $\rho$ is the surface density of galaxies brighter in apparent magnitude
than the fainter galaxy of the pair, $\theta$ is the pair separation and
$\beta$ an angular separation cut-off below which individual objects cannot
be resolved.
Woods et al. (1995) counted the numbers of $P\leq 0.05$ and 
$P\leq 0.10$ pairs on a ground-based CCD image with 869 galaxies of
$21<I<24$, but did not find a significant excess above the
random expectation, although Burkey et al. (1994) did find an excess when applying this technique to HST data of
similar depth.

We apply this method to our $K\leq 20$ galaxies,  
considering the Redeye and Magic data separately. 
Rather than sum $\pi\rho\alpha^2$ to estimate the area around each galaxy,
we place 50000 random points on the image areas and sum the number of points at
$\beta<\alpha<\theta$, thus taking into account the reduced areas around galaxies
near the edges. 
 We adopt $\beta=1.0$ arcsec (our catalog contains
no pairs of smaller separation), calculate $P$ for each pair of galaxies
and count all $P\leq 0.05$ and $P\leq 0.10$ pairs in $\Delta(\theta)=1$ arcsec bins of separation.

To estimate the number of pairs expected by chance, 100 
randomized datasets were generated, in which the galaxies retain their true magnitudes but are
randomly positioned within each field. An identical pair count analysis was performed on the randomized datsets and the results averaged, with the scatter between the simulations used to derive the $1\sigma$ noise errors of a single
random dataset.
\smallskip 
\noindent {\bf 6.2 Results}
\smallskip
Figure 7 shows histograms of the observed numbers of $P\leq 0.05$
and $P\leq 0.10$ pairs, with the random expectations and error bars.    
A significant excess above random of close pairs is seen only on the Redeye data, and
only at $2<\theta<3$ arcsec ($\sim 10$--16 $h^{-1}$ kpc at the mean
redshift). With a $P\leq 0.05$ threshold, we count 21 pairs of
$2<\theta<3$ arcsec on the Redeye fields compared to $9.76\pm 2.97$ expected by chance, a
$3.8\sigma$ significance excess of $11.24\pm 2.97$ pairs, or a factor of $2.15\pm 0.26$. With a $P\leq 0.10$ 
threshold, we count 27 pairs 
at this separation compared to $14.19\pm 3.67$ expected by chance, giving a 
slightly increased number of excess pairs, $12.81\pm 3.67$, but at a slightly 
lower significance of $3.5\sigma$.  There is no increase in the number of
excess pairs on moving the magnitude limit further faintward,
 presumably as a result of incompleteness at $K>20$. All the $P\leq 0.05$ pairs were examined by eye and found
to be genuine close pairs of galaxies rather than, for example, fragmented
 noise images.
 
The Redeye fields contain 529 galaxies at $K\leq 20$, so we estimate that $4.25\pm 1.12\%$ belong to 
$P\leq 0.05$ pairs in excess of the random expectation. Clustering with a $\theta^{-0.8}$ power-law of amplitude $\alpha$ at one arcsec would
increase the number of $2<\theta<3$ arcsec pairs above the random expectation by a factor of $1+0.48\alpha$. The observed  pair excess
indicates $\alpha=2.4\pm 0.6$, which extrapolates to 
$A=3.4\pm 0.9\times 10^{-3}$
at one degree. This exceeds the fitted $\omega(\theta)$ amplitude at $K\leq 20$ by a factor of $\sim 3.4/1.41=2.4$ (and $\sim 2.2\sigma$).
Hence there is some evidence that our data shows a similar enhancement
in the $\omega(\theta)$ amplitude at scales corresponding to $r<20$ $h^{-1}$ kpc
as the larger Infante et al. (1996) survey.

Some very close galaxy pairs will be
missed as the images are merged into a single detection. The close pair counts and $\omega(\theta)$  suggest that faint galaxy
pairs are only reliably split on our Redeye data if the projected separation
is exceeds 2 arcsec whereas on the Magic data a separation
of 3 arcsec may be required. 
If the true correlation function follows a $\theta^{-0.8}$ power-law 
at $\theta\leq 3$ arcsec, the number of $\theta<2$ arcsec pairs
will exceed the random expectation of 11.88 for the Redeye data by a factor of 
$0.96\alpha$. If we assume that all pairs were detected at $2<\theta<3$ arcsec, 
the value of $\alpha$ estimated at this separation
would indicate that 
there are a further $0.96\times 11.88\times 2.4\pm 0.6=27.4\pm 7.2$
excess pairs at $\theta<2$ arcsec, which are merged on our images. 
If this is the case, the true fraction of galaxies belonging to
$\theta<3$ arcsec pairs above the random expectation is
$\sim 14.6\pm 3.8\%$. As the excess of $2<\theta<3$ arcsec pairs above the random expectation exceeded that expected from the fitted $\omega(\theta)$ power-law by a factor $\sim 3.4/1.41=2.4$,  we estimate the fraction of galaxies in close pairs in excess of the larger scale $\omega(\theta)$ as 
$(1.4/2.4)\times 14.6\pm 3.8 = 8.5\pm 2.2\%$.

\bigskip
\noindent {\bf 7. Discussion}
\smallskip
On $K$-band data covering 
101.5 $\rm arcmin^2$ to $K\sim 20$, we detect galaxy clustering at the
$\sim 4\sigma$ level of significance. The
amplitude of this clustering appears to be a factor of $\sim 2$ greater than predicted by a PLE model with $r_0=5.9$ $h^{-1}$
Mpc and $\gamma=1.8$ for ellipticals, despite the good fit of this model 
 to the $\omega(\theta)$ scaling in the red and blue passbands (Roche et al. 1996a), and the $\omega(\theta)$ amplitude at $K\sim 15$--16 (from Baugh et al. 1996). We considered two possible explanations:

\noindent (i) The $\xi(r)$ of the early-type galaxies, especially prominent in $K$-limited samples, is 
significantly steeper than $\gamma=1.8$. 
Increasing the giant E/S0 galaxy $\xi(r)$ in our PLE model to  $r_0=8.25 h^{-1}$
Mpc with $\gamma=2.05$, as estimated by Guzzo et al. (1997), increased the
predicted  $K\leq 20$ galaxy $\omega(\theta)$ at  $\theta\sim 5$ arcsec more than enough  for consistency with our observations, provided that
galaxy clustering is approximately stable out to $z\sim 2$.
This steeper $\xi(r)$ would have less effect on the
$\omega(\theta)$ measured on larger scales on large $K$-band images 
(e.g. Baugh et al. 1996), or
on any blue and red limited surveys where the fraction of ellipticals will be much smaller, so should be consistent with these observations.

\noindent (ii) The trend towards more clustered environments for higher redshift radio galaxies might continue beyond $z\sim 0.5$, to reach $B_{rg}\sim 700$ at $z\sim 1.1$.
This would mean that $\sim 30\%$ of the $K\leq 20$ galaxies on our images belong to rich clusters associated with the radio galaxies,
and would increase the model B $\omega(\theta)$ amplitude
to at least the observed value, without any change in the assumed
clustering properties of field galaxies.

Although with this data we cannot firmly distinguish between these possibilities, there are a number of arguments favouring the first;

\noindent (i) We did not detect a significant cross-correlation between 
the radio galaxies on these fields and the surrounding  $K\leq 20$ galaxies,
obtaining upper limits which disfavour the Abell 1--2 clustering required to
significantly increase the galaxy $\omega(\theta)$ amplitude.

\noindent (ii) Ellingson and Green (1993) propose that radio sources in clusters undergo more rapid luminosity 
evolution out to $z\sim 0.6$ than field sources, but find that at higher redshifts the source number counts would constrain the evolution of field and cluster radio galaxies to be similar. If this is the case, the mean $B_{rg}$ 
at $z>0.6$ is likely to remain close to
the Abell class 0 value seen by Hill and Lilly (1991).

\noindent (iii) The deeper ($K\leq 21.5$) survey of Carlberg et al. (1997)
was not known to contain any radio galaxies, yet gave a relatively high $\omega(\theta)$
amplitude similar to that from our data, and even more discrepant from model B. 

\noindent (iv) When Carlberg et al. (1997) divided the $0.3\leq z\leq 0.9$ 
galaxies in their $K<20$ redshift survey at the median $U-K$ colour, the redder subsample 
galaxies gave a $\xi(r)$ of both a much higher amplitude and a steeper slope (closer to $\gamma=2.2$ than $\gamma=1.8$), than the bluer galaxies
 (which were consistent with $\gamma=1.8$).

It is also apparent that the high $\omega(\theta)$ amplitudes measured here
and by Carlberg et al. (1997) are more consistent with approximately 
stable ($\epsilon\sim 0$) small-scale galaxy clustering than with the rapid evolution ($\epsilon\sim 1.5$) suggested by some observers  (e.g. Shepherd et al. 1997). The low $\xi(r)$ amplitudes measured from $R$ or $I$ limited redshift
surveys (Le Fevre et al. 1996; Shepherd et al. 1997) would then be indicative of weak intrinsic clustering for very blue galaxies rather than a general rapid clustering evolution. Stable clustering is consistent with $\Omega \simeq 0.2$ simulations whereas $\epsilon\sim 1$ is expected for $\Omega=1$ models (Colin, 
Carlberg and Couchman 1997).

The Carlberg et al. (1997) $\omega(\theta)$ amplitude of $11.4\pm 1.6\times 10^{-4}$ exceeds even the model D prediction and may suggest
a levelling out of the scaling at 
$K>19.5$. This
contrasts with $\omega(\theta$) results obtained in the red-band, where 
at $R=26$ (Brainerd, Smail and Mould 1995) and
on the Hubble Deep Field at $R=28$ (Villumsen et al. 1997)
the $\omega(\theta)$ amplitude falls much further to
$\sim 2\times 10^{-4}$. 
This would suggest that the
galaxies with the reddest observer-frame $R-K$ colours, i.e. passively
evolving galaxies seen at $z>0.75$, are intrinsically very strongly clustered, perhaps even more so than local ellipticals.
 This might be expected from the models of Bagla (1987), which predict that at high redshifts ($z\sim 3$) the highest peaks in the density field would be much more clustered than the underlying mass
distribution. These peaks would be the first sites
for galaxy formation, and would form elliptical rather than spirals
(Balland, Silk abd Schaeffer 1997). Hence a `first generation' of early-type galaxies,
completing its star-formation prior to $z\sim 3$ and  
becoming very red in $R-K$ at $z\sim 1$--2, should be more strongly clustered
than other galaxies.

A population of very red galaxies begins to enter surveys at $K\sim 18.5$
(Djorgovski et al. 1995 ; Cowie et al. 1996; Moustakos et al. 1997), so if strongly clustered may  first produce an impact on $\omega(\theta)$ -- an upturn relative to the standard models -- at the depths of this
survey. However, deep imaging of a much larger area in $K$ and $R$ will be needed to confirm whether these high redshift red galaxies are really more clustered than local ellipticals.

The mean clustering environment of $z\sim 1$ radio galaxies
remains undetermined, except that a
mean environment richer than Abell class 1 appears to be disfavoured.
However, at a given redshift the environments of individual radio galaxies may differ greatly (see Hill and Lilly 1991; Yee and Green 1987).  
Deltorn et al. (1997) have confirmed by spectroscopy that at least one radio galaxy
(3CR184) in this redshift range is in a rich (Abell 2) cluster. 
Bowes and Smail (1997) detected gravitational lensing centred on another radio galaxy, 3C336, at $z=0.927$, indicating the presence of a massive cluster, but could only set upper limits on the clustering around 7 other
3C sources at $z\sim 1$.
 Allington-Smith et al. (1982) had previously imaged areas centred on 8 of the 17 galaxies in our sample
with a red-band CCD and found marginal (0.90--0.99
confidence) evidence of surface density enhancements around four (6C0822+39,
6C1129+37, 6C1217+36 and 6C1256+36), and for  
2 of the radio galaxies, 6C1011+36 and  6C1256+36, our fitted 
radiogalaxy-galaxy cross-correlation amplitudes are as high as $\sim 0.01$ (although with large errors), and the $\omega(\theta)$ amplitudes on these fields are also higher than average (2.4--3.2$\times 10^{-3}$), and consistent with Abell 2 clustering around these 2 galaxies only. 

 At these high redshifts, it is likely that further
data such as observations in more than one passband to colour-select galaxies
near the radio galaxy redshift, multi-fibre spectroscopy, or the detection of lensing or extended X-ray sources will be needed to confirm the existence of clusters.
 Considering the observations made so far and the likelihood that radio galaxy evolution is less dependent on cluster environment at $z>0.5$ (Yee and Ellingson 1993), it is most plausible that these 17 radio galaxies have a $B_{rg}$ distribution similar to that of radio galaxies at $z\sim 0.5$
(Hill and Lilly 1991), with one or two in Abell 1-2 clusters but the great majority  in field or Abell 0 environments.

Finally, on the Redeye data, we find a significant  excess of close ($2\leq \theta\leq 3$ arcsec) pairs
of galaxies above the random expectation, and above an inwards extrapolation of the fitted
$\omega(\theta)$ power-law. This close pair excess is similar to
size (a factor of $\sim 2$ above random) and in the 
corresponding range of physical separations ($r< 20 h^{-1}$ kpc)
to that reported 
by Yee and Ellingson (1995), Infante et al. (1996) and Patton et al. (1997)
on red-band data. 
No close-pair excess was seen on the Magic fields, which suggests that close
pair counts depend critically on the resolution of the data. This might
explain the difference between the results of Burkey at al. (1994) and 
Woods et al. (1995) and the wide variation in other close-pair
studies (see e.g. Neuschaefer et al. 1997).

After applying a correction for the number of close pairs missed due 
to image merging,
we estimated that $14.6\pm 3.8\%$ of the $K\leq 20$ galaxies on the Redeye fields belonged to close pairs in excess of the expectation for a random distribution, and that $8.5\pm 2.2\%$ belonged to pairs in excess of the
expectation from the fitted
$\omega(\theta)$. We estimate the enhancement in the number of close pairs above the random 
expectation as a factor of $2.15\pm 0.26$ compared with 1.637 seen by Infante et al.
(1996) at $z_{mean}=0.35$. If the merger rate increases with redshift as
$(1+z)^m$ and our sample has $z_{mean}\simeq 1.09$, $(2.09/1.35)^m
\simeq (2.15\pm 0.26)/1.637$ giving a very rough
estimate of $m\simeq 1.35\pm 0.50$. Comparing with 
 a local estimate of $4.5\%$ for the
fraction of galaxies at projected separations of $r<20 h^{-1}$ kpc
(Yee and Ellingson 1995), our results suggests that the galaxy merger rate at $z_{mean}\simeq 1.09$ is $\sim 14.6/4.5\sim 3.2$ times its local value, 
so that $2.09^m\simeq (14.6\pm 3.8)/4.5$ and therefore $m\simeq 1.60\pm 0.36$.
 
 As a $K$-band sample will not be biased towards starbursting objects, this 
evolution cannot 
be due to mergers
producing more luminous starbursts at earlier epochs, but must reflect a real increase in the merger frequency with redshift. The 
increase appears relatively modest compared to the $m\simeq 4$ predicted by some
$\Omega=1$ models (Carlberg, Pritchet and Infante 1994), but does seem consistent with
the surprisingly low Neuschaefer et al. (1997) estimate of $m\sim 1.2\pm 0.4$ from similarly deep
($I\leq 25$) HST data, and the $m\simeq 2$ predicted by low $\Omega$ models. 
\bigskip
\noindent {\bf 8. Summary of Main Conclusions} 
\smallskip
(i) Significant galaxy clustering was detected on ground based $K$-band images of 17 fields (totalling
101.5 $\rm arcmin^2$)  containing $z\sim 1$ radio galaxies. Fitting a $\theta^{-0.8}$ power law, the $\omega(\theta)$  amplitude at one degree was estimated as
 $1.33\pm 0.32 \times 10^{-3}$ at $K\leq 20$.

(ii) The $\omega(\theta)$ amplitudes at $K=20$ are consistent with that estimated by Carlberg et al. (1997) at $K=21.5$ but higher than predicted by the Roche et al. (1996) PLE model, known to fit the faint galaxy clustering in
blue and red passbands. These results remain consistent with a PLE model if
the clustering of giant ellipticals is similar to the  
Guzzo et al. (1997) estimate ($r_0\simeq 8.35$ $h^{-1}$ Mpc, $\gamma\simeq 2.05$) and galaxy clustering is stable out to $z\sim 2$.

(iii) We did not detect a significant cross-correlation between the $z\sim 1$ radio galaxies and the surrounding galaxies and obtained a
$1\sigma$ upper limit of $B_{rg}= 587$, consistent with the same mean clustering environment (Abell class 0) as that of radio galaxies at $z\sim 0.5$. This would argue against 
(but not firmly exclude) an Abell class 1--2 or richer
environment. As Abell class 0 clustering would have little effect on the galaxy 
$\omega(\theta)$, a steep $\xi(r)$ for E/S0 galaxies remains the
most likely explanation for its high amplitude.

(iv)  On the Redeye fields, the number of galaxy-galaxy pairs of $2\leq \theta\leq 3$  separation exceeds the random expectation by a factor 
of $2.15\pm 0.28$, more than expected from the fitted angular correlation function.
This close pair excess is comparable to that previously 
reported on $R$-band data (Infante et al. 1996), and suggests that the merger
rate at $z\sim 1$ is about three times the local rate.  
\bigskip
{\bf Acknowledgements}

Nathan Roche acknowledges the support of a PPARC research associateship.
 We thank Garrett Cotter for help with the observations.
\smallskip
\bigskip 
{\bf References}
\bigskip

\hangindent=2pc \hangafter=1 Allington-Smith,~J.R., Perryman,~M., Longair,~M.S.,
Gunn,~J.E. and Westphal,~J.A., 1982.
 {\it Mon. Not. R. astr. Soc.\/}, {\bf 201}, 331.

\hangindent=2pc \hangafter=1 Balland,~C., Silk,~J. and Schaeffer,~R., 1997.  
{\it Astrophys. J.\/}, in press.

\hangindent=2pc \hangafter=1 Bagla,~J.S., 1997.
 {\it Mon. Not. R. astr. Soc.\/}, submitted.

\hangindent=2pc \hangafter=1 Baugh,~C.M., Gardner,~J.P., Frenk,~C.S. and
Sharples,~R.M., 1996.  
{\it Mon. Not. R. astr. Soc.\/}, {\bf 283}, P15.

\hangindent=2pc \hangafter=1 Bowes,~R.G., and Smail,~I., 1997.  
{\it Mon. Not. R. astr. Soc.\/}, {\bf 290}, 292.

\hangindent=2pc \hangafter=1 Brainerd,~T.G, Smail,~I. and Mould,~J., 1995.  
{\it Mon. Not. R. astr. Soc.\/}, {\bf 275}, 781.

\hangindent=2pc \hangafter=1 Bruzual,~G. and Charlot,~S., 1993.  
{\it Astrophys. J.\/}, {\bf 405}, 538.

\hangindent=2pc \hangafter=1 Burkey,~J.M., Keel,~W.C., Windhorst,~R.A. and
Franklin,~B.E., 1994.  
{\it Astrophys. J.\/}, {\bf 429}, L13.

\hangindent=2pc \hangafter=1 Carlberg,~R.G, Pritchet,~C.J. and
 Infante,~L., 1994.  
{\it Astrophys. J.\/}, {\bf 435}, 540.

\hangindent=2pc \hangafter=1 Carlberg,~R.G., Cowie,~L.L., Songaila,~A.
and Hu,~E.M., 1997. {\it Astrophys. J.\/}, {\bf 484}, 538.

\hangindent=2pc \hangafter=1 Colin,~P., Carlberg,~R.G. and Couchman,~H.M.P., 1997. {\it Astrophys. J.\/}, submitted.

\hangindent=2pc \hangafter=1 Cowie,~L.L., Hu,~E.M. and Songaila,~A.
1995.  {\it Nature}, {\bf 377}, 603.

\hangindent=2pc \hangafter=1 Cowie,~L.L.,  Songaila,~A. and Hu,~E.M.,
1996.  {\it Astron. J. \/}, {\bf 112}, 839.

\hangindent=2pc \hangafter=1 Deltorn,~J.-M., Le F\`evre,~O., Crampton,~D.
and Dickinson,~M., 1997. {\it Astrophys. J.\/}, {\bf 483}, L21.
 
\hangindent=2pc \hangafter=1 Djorgovski,~S.G., et al., 1995.
{\it Astrophys. J.\/},  {\bf 438 }, L13. 

\hangindent=2pc \hangafter=1 Eales,~S., Rawlings,~S., Law-Green,~D., 
Cotter,~G. and Lacy,~M., 1997. {\it Mon. Not. R. astr. Soc.\/}, submitted.

\hangindent=2pc \hangafter=1 Gardener,~J.P., Cowie,~L.L. and Wainscoat,~R.J.,
1994. {\it Astrophys. J.\/},  {\bf 415}, L9. 

\hangindent=2pc \hangafter=1 Glazebrook,~K., Peacock,~J., Collins,~C. and Miller,~L., 1994. {\it Mon. Not. R. astr. Soc.\/}, {\bf 266}, 65.

\hangindent=2pc \hangafter=1 Guzzo,~L., Strauss,~M.A., Fisher,~K., 
Giovanelli,~R. and Haynes,~M., 1997.
{\it Astrophys. J.\/},  {\bf 489}, in press.

\hangindent=2pc \hangafter=1 Hill,~G.J. and Lilly,~S.J., 1991.
{\it Astrophys. J.\/},  {\bf 367}, 1. 

\hangindent=2pc \hangafter=1 Infante,~L., de Mello,~D,F., and Menanteau,~F., 1996. {\it Astrophys. J.\/},  {\bf 469}, L85. 

\hangindent=2pc \hangafter=1 Le F\`evre,~O., Hudon,~D., Lilly,~S.J., Crampton,~D., Hammer,~F. and Tresse,~L. 1996. {\it Astrophys. J.\/}, 
 {\bf 461}, 534. 

\hangindent=2pc \hangafter=1 Longair,~M.S. and Seldner,~M.,
1979. {\it Mon. Not. R. astr. Soc.\/}, {\bf 189}, 433.

\hangindent=2pc \hangafter=1 Loveday,~J., Maddox,~S.J., Efstathiou,~G. and
Peterson,~B.A. 1995. {\it Astrophys. J.\/}, {\bf 442}, 457. 

\hangindent=2pc \hangafter=1 Moustakas,~L., Davis,~M., Graham,~J.R,
Silk,~J., Peterson,~B.A. and Yoshii,~Y., 1997.
{\it Astrophys. J.\/}, {\bf 475}, p.445.

\hangindent=2pc \hangafter=1 Neuschaefer,~L.W., Im,~M., Ratnatunga,~K.U., Griffiths,~R.E. and Casertano,~S., 1997. {\it Astrophys. J.\/}, {\bf 480}, 59.

\hangindent=2pc \hangafter=1 Patton,~D.R., Pritchet,~C.J., Yee,H.K.C.,
Ellingson,~E. and Carlberg,~R.G., 1997.
{\it Astrophys. J.\/},  {\bf 475}, 29.

\hangindent=2pc \hangafter=1 Prestage,~R. and Peacock,~J.A., 1988. {\it Mon. Not. R. astr. Soc.\/}, {\bf 230}, 131.

\hangindent=2pc \hangafter=1 Prestage,~R. and Peacock,~J.A., 1989. {\it Mon. Not. R. astr. Soc.\/}, {\bf 236}, 959.

\hangindent=2pc \hangafter=1 Roche,~N., Shanks,~T., Metcalfe,~N. and
Fong,~R., 1996a.
 {\it Mon. Not. R. astr. Soc.\/}, {\bf 280}, 397.

\hangindent=2pc \hangafter=1 Roche,~N., Ratnatunga,~K., Griffiths,~R.E,
Im,~M. and Neuschaefer,~L., 1996b
 {\it Mon. Not. R. astr. Soc.\/}, {\bf 282}, 1247.

\hangindent=2pc \hangafter=1 Roche,~N., Ratnatunga,~K., Griffiths,~R.E.,
and Im,~M., 1997a.
 {\it Mon. Not. R. astr. Soc.\/}, submitted.

\hangindent=2pc \hangafter=1 Roche,~N., Eales,~S., and Rawlings,~S. 1997b.
 {\it Mon. Not. R. astr. Soc.\/}, submitted.

\hangindent=2pc \hangafter=1 Shepherd,~C.W., Carlberg,~R.G.. and
Yee,~H.K.C., 1997. {\it Astrophys. J.\/},  {\bf 479}, 82.

\hangindent=2pc \hangafter=1 Steidel,~C.C., Giavalisco,~M., Dickinson,~M.,
and Adelberger,~K.L., 1996. {\it Astron. J.\/}, {\bf 112}, 352.

\hangindent=2pc \hangafter=1 Villumsen,~J.V., Freudling,~W. and
da Costa,~L. , 1997. {\it Astrophys. J.\/},  {\bf 481}, 578.

\hangindent=2pc \hangafter=1 Wainscoat,~R.J., and Cowie,~L., 1992.
{\it Astron. J.\/}, {\bf 103}, 323.

\hangindent=2pc \hangafter=1 Woods,~D., Fahlman,~G and Richer,~H., 1995.
{\it Astrophys. J.\/},  {\bf 454}, 32.

\hangindent=2pc \hangafter=1 Yates,~M.G., Miller,~L. and Peacock,~J.A.,
1989. {\it Mon. Not. R. astr. Soc.\/}, {\bf 240}, 129.

\hangindent=2pc \hangafter=1 Yee,~H.K.C. and Ellingson,~E., 1993.
{\it Astrophys. J.\/},  {\bf 411}, 43. 

\hangindent=2pc \hangafter=1 Yee,~H.K.C. and Ellingson,~E., 1995.
{\it Astrophys. J.\/},  {\bf 445}, 37. 

\hangindent=2pc \hangafter=1 Yee,~H.K.C. and Green,~R.F., 1987.
{\it Astrophys. J.\/},  {\bf 319}, 28.

\bigskip
\noindent {\bf Table Captions}
\bigskip
\noindent{\bf Table 1.} The co-ordinates of the 17 radio galaxies, their 
$K$-band magnitudes as measured from this data, spectroscopic redshifts where
available, and the ($1.5\sigma$) 
detection thresholds of the images. The two bracketed redshifts are 
very approximate estimates derived from the $K-z$ relation.
\medskip
\noindent{\bf Table 2.} The total number of galaxies and the fitted angular
correlation function amplitudes (at $\theta=1$ degree assuming a $\theta^{-0.8}$ power-law) for the Redeye and Magic fields and for the combined dataset.
\medskip
\noindent{\bf Table 3.} Our estimates of the three quantities discussed in 
Section 5.3 for the each radio galaxy. Bracketed values are those
calculated using the bracketed estimated redshifts in Table 1.
\bigskip
\noindent{\bf Figure Captions}
\bigskip
\noindent{\bf Figure 1.} Differential galaxy number counts in $\Delta(K)=0.5$ mag intervals, from this data and the previous $K$-band surveys of Moustakas 
et al. (1997), Glazebrook et al. (1994), Gardner, Cowie and Wainscoat (1995)
and Djorgovski et al. (1995), compared with $q_0=0.05$ models with no evolution
(dotted) and pure luminosity evolution (dashed).
\medskip
\noindent{\bf Figure 2.} The mean angular correlation functions, $\omega(\theta)$, of the galaxies on the 12 Redeye and 5 Magic fields, with field-to-field errors. The dotted lines show the best-fitting functions
of the form `$A(\theta^{-0.8}-36.0)$' for the Redeye fields and `$A(\theta^{-0.8}-22.0)$'for the Magic fields, with amplitudes $A$ as listed
in Table 2. 
\medskip
\noindent{\bf Figure 3.} The $\omega(\theta)$ amplitude from all 17  fields combined, and from   Baugh et al. (1996) and Carlberg et al. (1997), as a function of $K$-band magnitude limit. The dotted line shows the non-evolving model A, the solid line a PLE model (model B) with stable clustering ($\epsilon=0$), the dashed line a PLE model
with $\epsilon=1.2$ clustering evolution (model C) and the dot-dash line the clustering measured at small separations of $\theta\sim  5$ arcsec for the PLE model with increased and steepened clustering for E/S0 galaxies (model D). 
\medskip
\noindent{\bf Figure 4.} Galaxy redshift distribution predicted by our luminosity evolution model at $K\leq 20$, for all types of galaxy (dotted) and for E/S0 galaxies only (dashed).
\medskip
\noindent{\bf Figure 5.} The angular cross-correlation between radio galaxies and the surrounding galaxies, averaged for all 17 radio galaxies with field-to-field errors  and the best-fitting function of the form 
`$A(\theta^{-0.8}-36.2)$'
\medskip
\noindent{\bf Figure 6.} (a) The predicted amplitude of the angular radiogalaxy-galaxy cross-correlation for 
$K\leq 20$ galaxies on these fields, as a function of  $B_{rg}$. The triangles (top) show the values of $B_{rg}$ corresponding to the clustering of field E/S0 galaxies and that in Abell class
0, 1 and 2 clusters. The circular symbol, arbitrarily placed at
$B_{rg}=400$, indicates our observed cross-correlation amplitude. (b)
The $\omega(\theta)$ amplitude expected for $K\leq 20$ galaxies on these
fields, with field galaxy clustering as given by PLE model B, as a 
function of $B_{rg}$.  The circular symbol, arbitrarily placed at
$B_{rg}=400$, indicates our observed $\omega(\theta)$ amplitude.
\medskip
\noindent{\bf Figure 7.} Histograms showing the observed numbers of
 close galaxy-galaxy pairs on the Redeye and Magic fields, with probability
thresholds of $P<0.05$ and $P<0.10$. The dotted lines show the pair counts from 
simulated random distributions of galaxies, the error bars indicate
$1\sigma$ errors from the simulations.
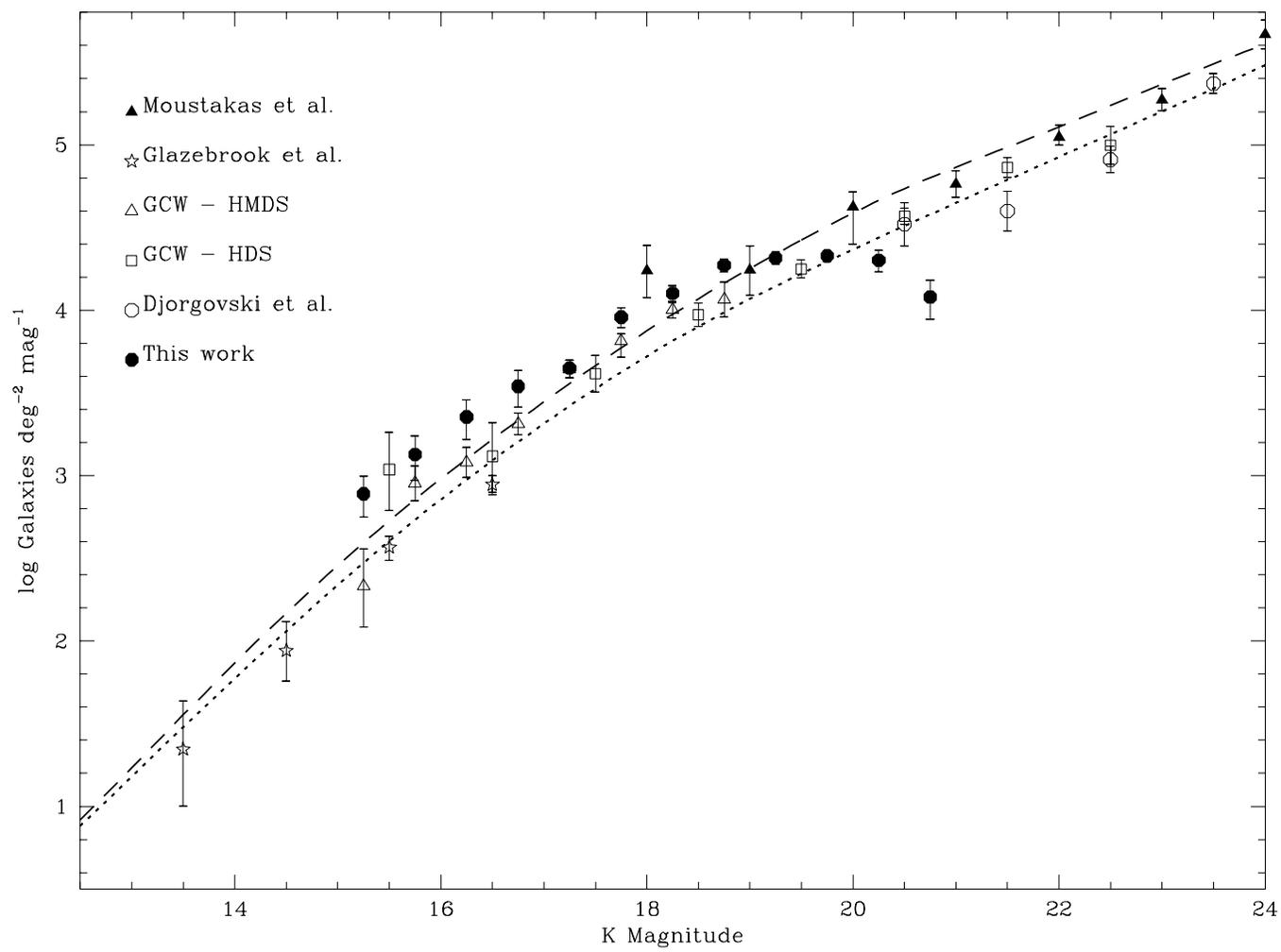
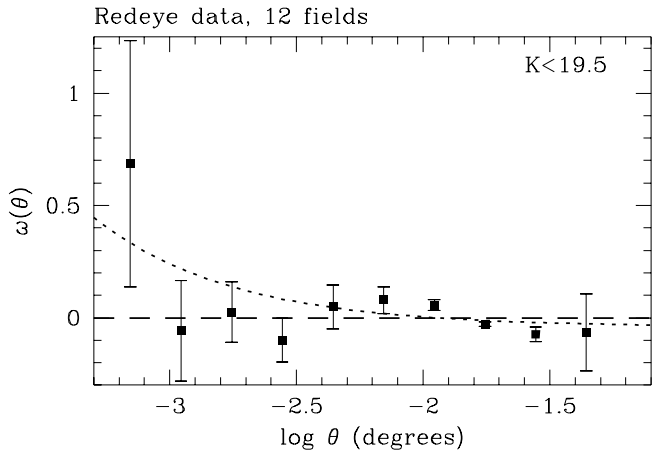
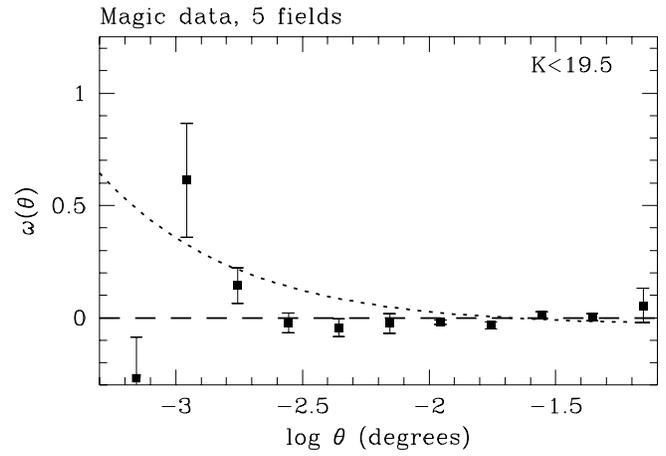
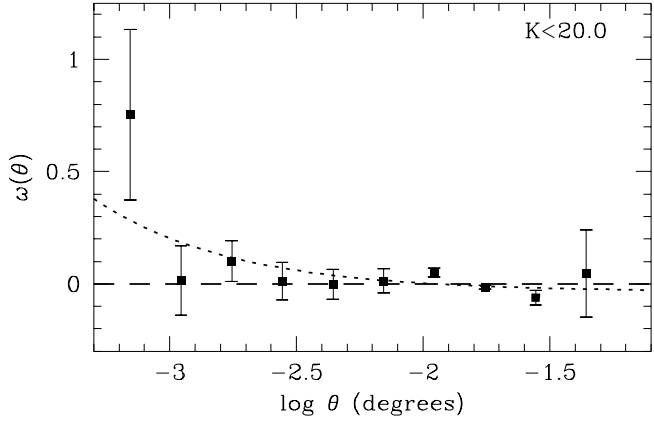
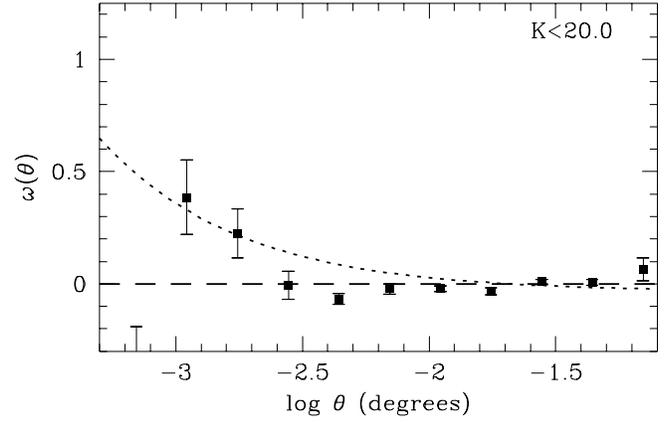
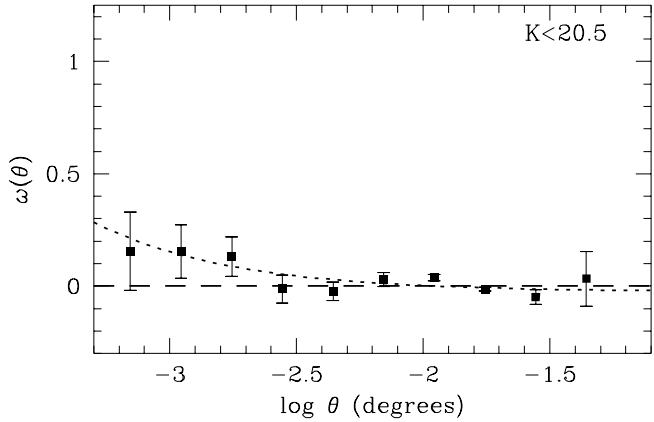
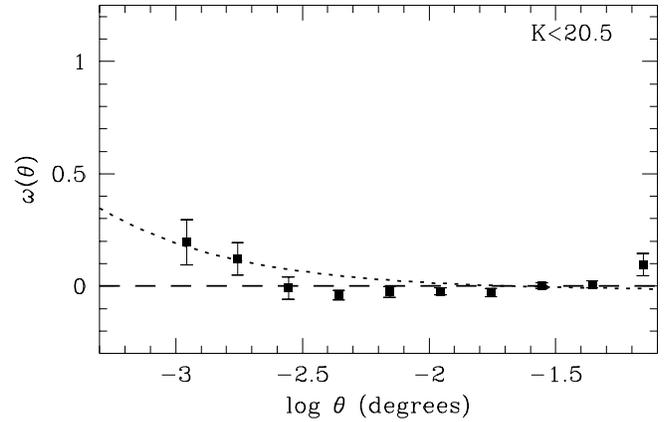
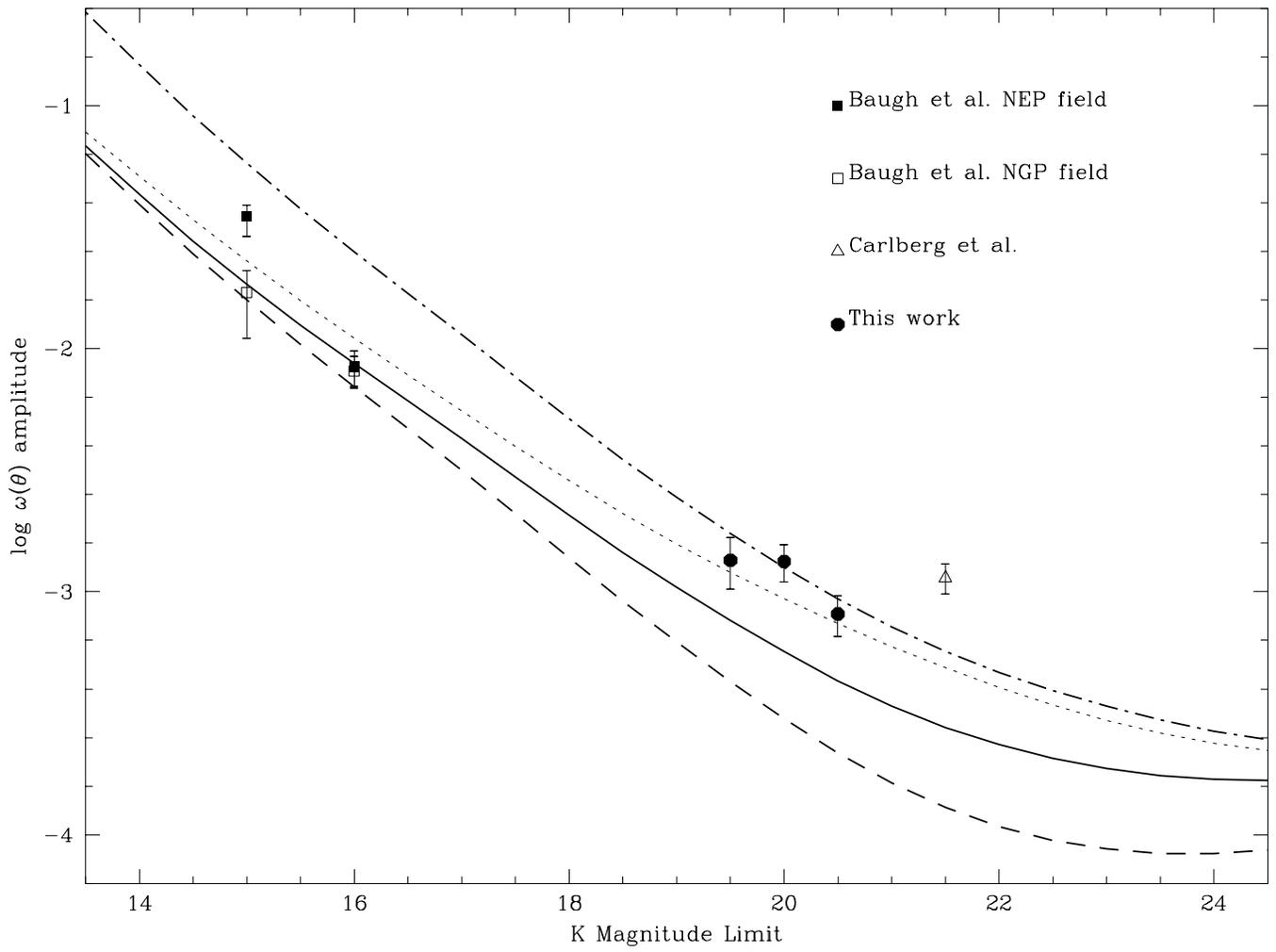
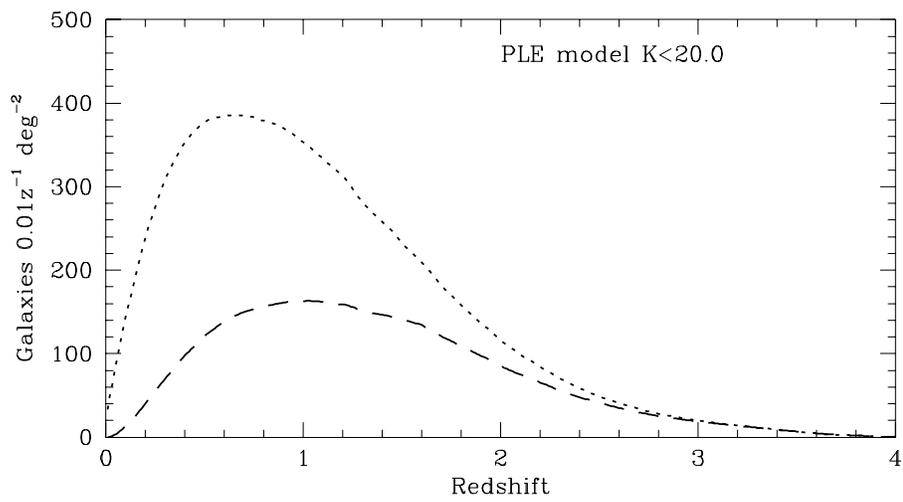
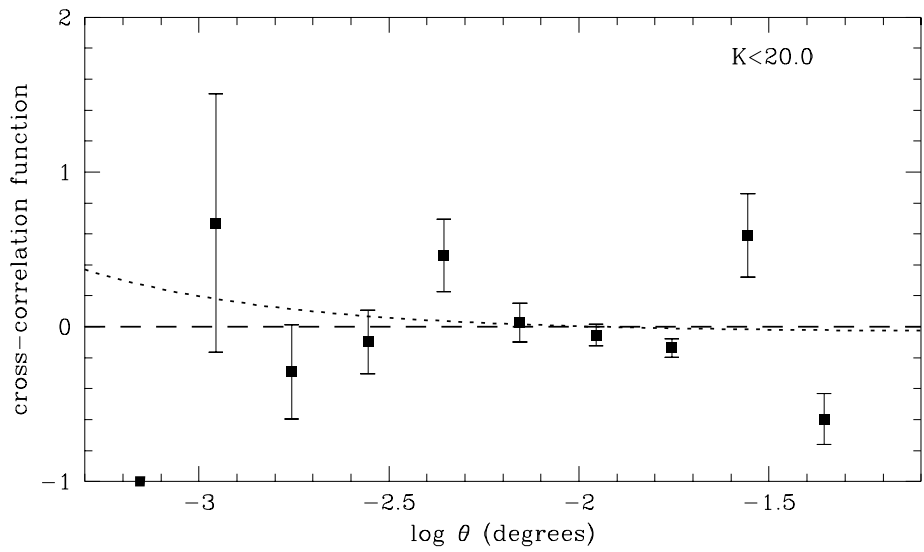
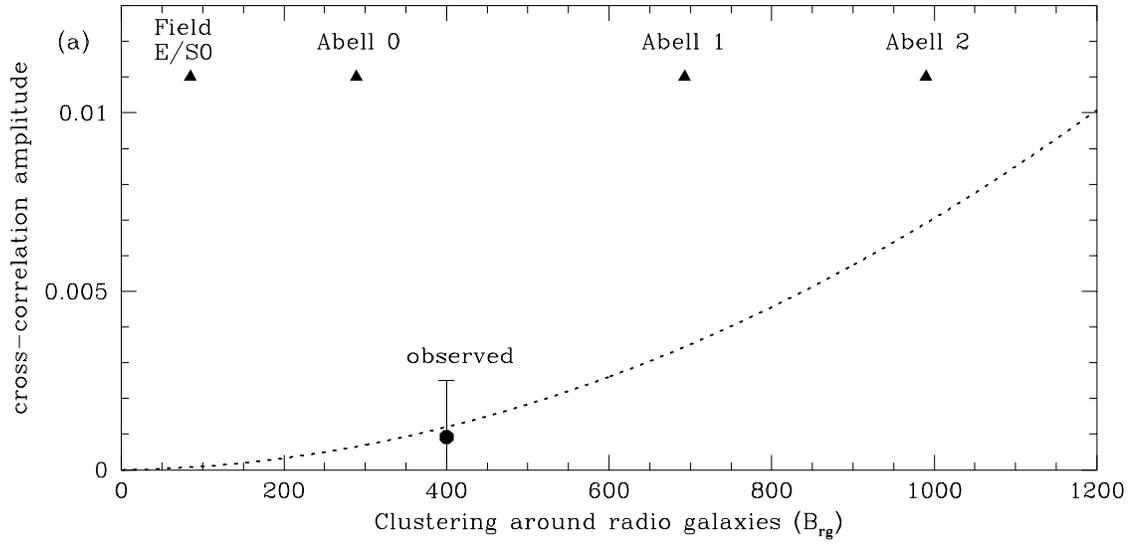
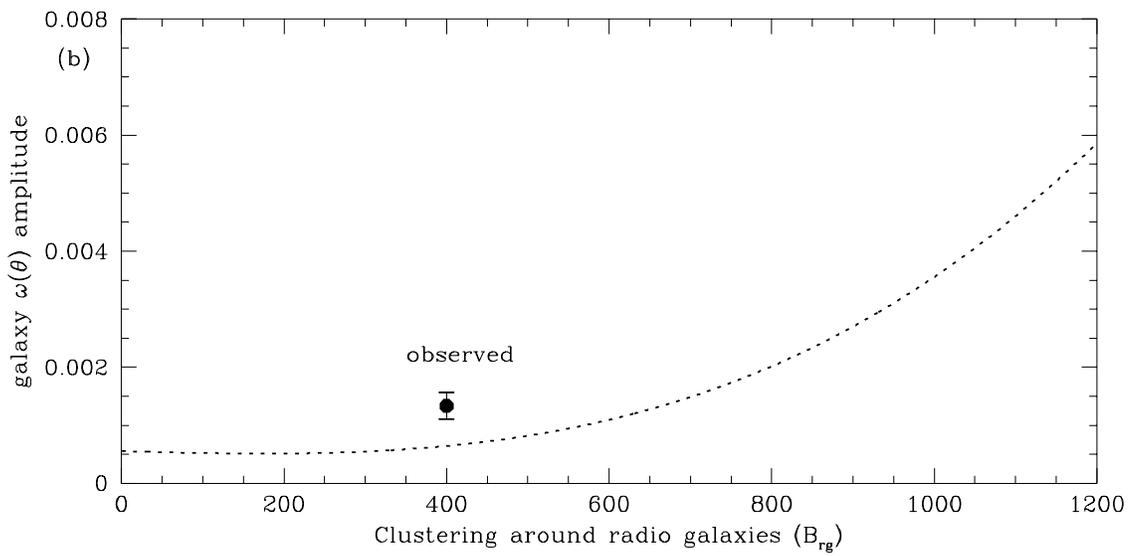
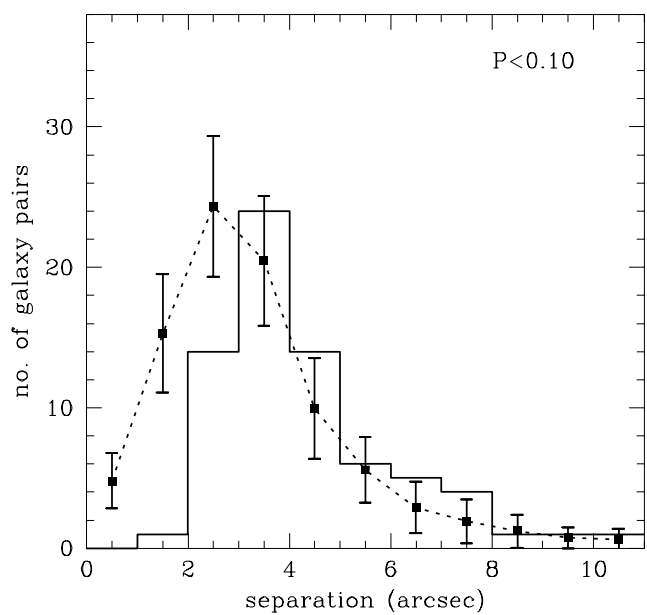
\end